\documentclass[%
aip,
jap,
reprint, 
floatfix,
noshowpacs,
noshowkeys,
preprintnumbers
]{revtex4-2}
\usepackage{mathptmx}
\usepackage{graphicx}
\usepackage{etoolbox}
\usepackage{dcolumn}
\usepackage{float}
\usepackage{txfonts}
\usepackage[english]{babel}
\newlength{\figsize}
\newlength{\subfigsize}
\makeatletter
\newcommand{\manuallabel}[2]{\def\@currentlabel{#2}\label{#1}}
\makeatother

\begin{document}
\title{Strong field confinement in air and enhanced optical {Kerr} nonlinearity of high-index star-shaped pore membrane metasurfaces}
\author{Andrey V. Panov}
\affiliation{
Institute of Automation and Control Processes,
Far Eastern Branch of Russian Academy of Sciences,
5, Radio st., Vladivostok, 690041, Russia}
\email{panov@iacp.dvo.ru}
\begin{abstract}
The optical Kerr effect in high-refractive-index membrane metasurfaces patterned with a square lattice of five-pointed star-shaped pores is investigated numerically using gallium phosphide as a model material.
The star geometry supports resonances with pronounced electromagnetic energy confinement, including an anapole state.
The sharp tips of the star-shaped pores concentrate the electric energy density in air, exceeding that of Mie voids by a factor of seven.
The effective second-order refractive index is enhanced by up to three orders of magnitude relative to bulk GaP, with the largest values obtained near the anapole resonances.
The nonlinear response remains stable under polarization rotation and random misalignment of the nanoholes, indicating tolerance to fabrication imperfections.
These results establish star-shaped pore lattices in freestanding membranes as a promising platform for nonlinear optics exploiting strong field confinement.
\end{abstract}
\maketitle



The enhancement of optical nonlinearities in metasurfaces has become a major research theme, driven by the need for efficient frequency conversion, all-optical switching, and quantum light generation.
The nonlinear response of a metasurface can be substantially boosted by exploiting resonant modes that localize the fundamental field within the nonlinear material.
Resonances with strong electromagnetic field enhancement have proven particularly effective in this context, since they allow the nonlinear interaction to be amplified well beyond the level achievable in an unpatterned film.
This effect is governed by the field confinement and by the spatial overlap of the resonant mode with the nonlinear material, and it can be further increased by engineering the geometry of the unit cell.


Membrane metasurfaces (freestanding perforated slabs) that eliminate substrate-induced leakage and restore vertical symmetry have recently emerged as an ideal platform for maximizing optical field confinement, since they support high-quality-factor guided-mode resonances or quasi-bound states in the continuum (quasi-BIC) with fields strongly localized in the high-index material \cite{Yang20, Zheng23, Xu22}.
For instance, in these structures, the periodic hole array machined through a high-index slab hosts electric- and magnetic-dipole Mie resonances in the dielectric domains between the holes \cite{Yang20}, or supports collective strong coupling between cavity photons and molecular vibrations hosted in additional air voids \cite{Adi24}, as well as giant enhancement of the photoluminescence emission of a WS$_2$ monolayer placed on top of the membrane \cite{Deng26}.
In such systems, the optical Kerr effect (OKE), which manifests itself as an intensity-dependent refractive index $n_2$, has been measured experimentally with enhancement factors reaching several times that of the bulk value, underscoring the potential of membrane geometries for nonlinear applications \cite{Franceschini26}.
Membrane metasurfaces exhibiting the multipoint Kerker effect with quasi-BIC resonances were proposed for spectral and spatial selectivity of a deflected beam and its low-threshold modulation \cite{Fan25}.
In addition, several representative examples of membrane metasurfaces can be mentioned: a perforated lithium-niobate membrane generated second harmonics with an efficiency of $2\times10^{-4}$ \cite{Qu22}; silicon dimer-hole membranes upconverted infrared images to the visible through third-harmonic generation \cite{Zheng23,Pu23}; and doubly resonant silicon membranes strongly enhanced four-wave mixing between near- and short-wave-infrared beams \cite{Xu22}.

A complementary approach to field confinement is offered by Mie voids---spherical or truncated conical air cavities embedded in a high-index medium.
In contrast to conventional Mie resonators, where the field is concentrated inside the high-index material, Mie voids localize the electromagnetic field within the low-index void region, making the field maxima accessible to nearby emitters or nonlinear media placed inside the cavity \cite{Hentschel23, Ryabkov26}.
This inverted geometry preserves the multipolar resonance structure while enabling field enhancement in air, which is particularly advantageous for coupling with two-dimensional materials, molecular vibrations, and quantum emitters \cite{Sarbajna26}.


In this work, the OKE in high-index membrane metasurfaces patterned with star-shaped pores is investigated numerically.
The five-pointed star geometry supports several resonances with distinct field distributions, resulting in large enhancements of the OKE.
Moreover, the sharp tips of the star-shaped pores act as efficient field concentrators, leading to a pronounced enhancement of the electric energy density in the void regions.

The numerical procedure for retrieving the effective optical Kerr nonlinearity of the metasurface is based on the method introduced in Ref.~\cite{Panov18}.
Three-dimensional finite-difference time-domain (FDTD) simulations of a Gaussian beam propagating through the nonlinear sample are performed.
The phase shift of the transmitted beam is calculated at several light intensities $I$, which allows the effective second-order nonlinear refractive index $n_2$ to be evaluated from
$ n=n_0+n_2 I$,
where $n_0$ is the linear refractive index.
The nonlinear refractive index is estimated at multiple locations on the transmitted beam axis, giving the mean value and the standard deviation of $n_2$.
The effective nonlinear refractive index represents the ensemble-averaged optical response of the metasurface.

Details of the FDTD simulations are described in Refs.~\cite{Panov22,Panov23}.
The computational domain is $2.8\times 2.8\times 15$~$\mu$m with a spatial resolution of 3~nm.
The simulations are performed at the wavelength $\lambda=532$~nm.
Gallium phosphide (GaP) is selected as the metasurface material because of its high refractive index $n_{0\,\mathrm{in}}=3.49$ at this wavelength and negligible extinction coefficient \cite{Aspnes83}.
The thickness of the free-standing GaP membrane in vacuum is fixed at $h=100$~nm.
The effective second-order nonlinear refractive index of the metasurface is compared with that of bulk GaP and is presented as a unitless enhancement factor $n_{2{\mathrm{eff}}}/n_{2{\mathrm{bulk}}}$.
The multipole decomposition is based on the approximate formulas from Ref.~\cite{Alaee18}.
The linear refractive index $n_{0\,\mathrm{in}}$ is assumed to be constant over the whole wavelength range in these calculations.

\begin{figure}
{\centering\includegraphics[width=\figsize]{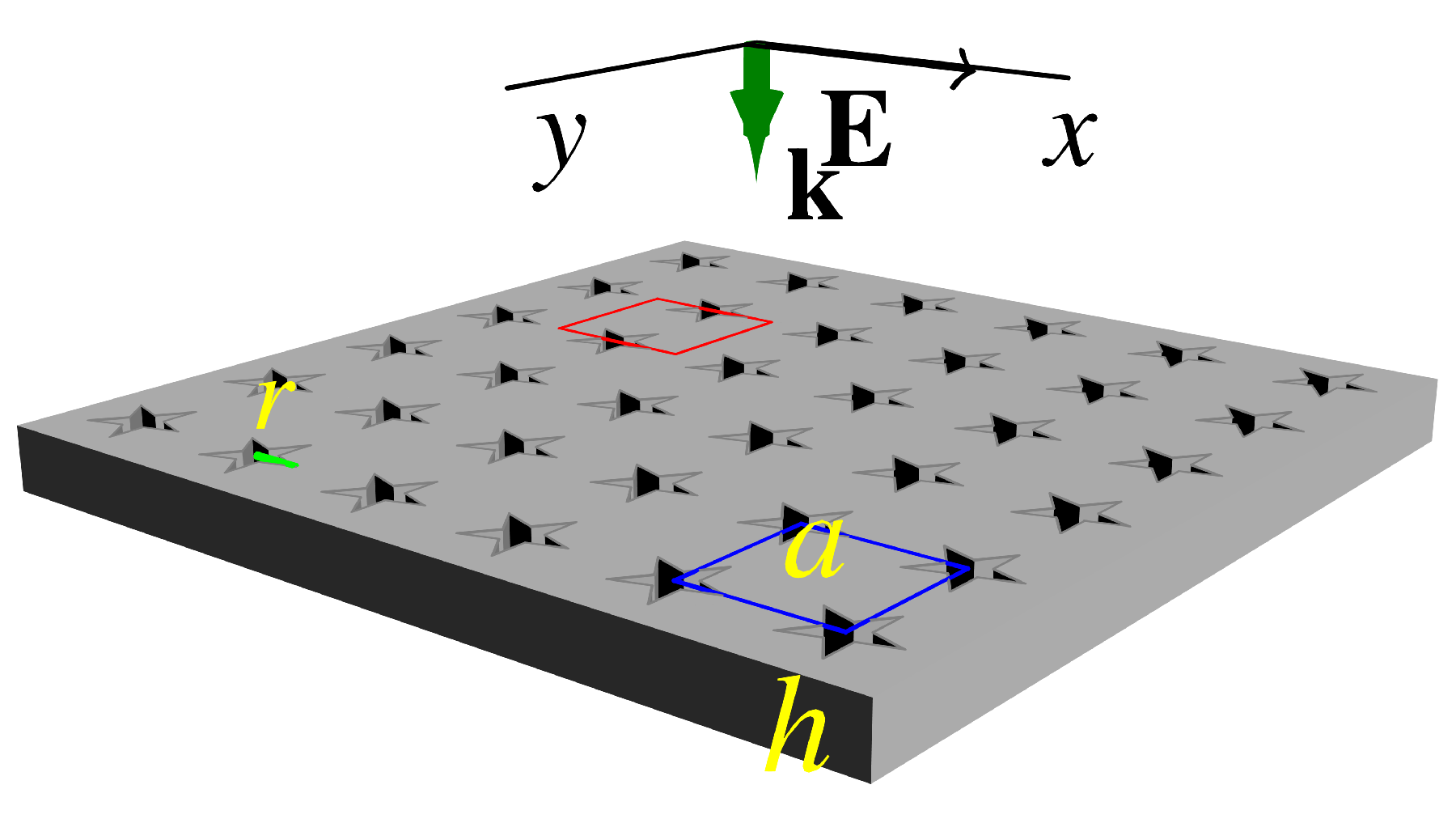}\par} 
\caption{\label{star_pores}  
Schematic of the simulated metasurface comprising a square lattice array of five-pointed star-shaped nanoholes in a high refractive index GaP membrane. 
The Gaussian beam is incident normally on the metasurface.}
\end{figure}

A schematic representation of the simulated metasurface is shown in Fig.~\ref{star_pores}.
Here $r$ is the radius of the circumscribed circle of the five-pointed star-shaped hole, $a$ is the square lattice constant, with the lattice element outlined by the blue line, and $h$ is the membrane thickness.
The red square illustrates another type of the lattice element utilized for the multipole analysis.
The linearly polarized Gaussian beam is normally incident on the metasurface.
Except where otherwise noted, the electric field of the incident beam is polarized along the $x$-axis.

\begin{figure}[tbh!]
{\centering
{
$a=240$~nm, $r=104$~nm
\par}
\includegraphics[height=\subfigsize]{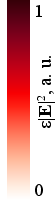}%
\includegraphics[width=\subfigsize]{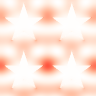}%
\hspace{1em}
\includegraphics[width=\subfigsize]{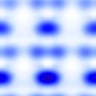}%
\includegraphics[height=\subfigsize]{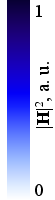}%
\\
{
$a=250$~nm, $r=122$~nm
\par}
\includegraphics[width=\subfigsize]{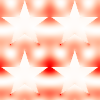}%
\hspace{1em}
\includegraphics[width=\subfigsize]{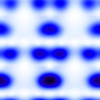}%
\\
{
$a=290$~nm, $r=142$~nm
\par}
\includegraphics[width=\subfigsize]{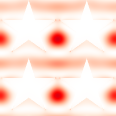}%
\hspace{1em}
\includegraphics[width=\subfigsize]{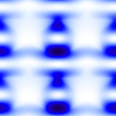}%
\\
{
$a=320$~nm, $r=117$~nm
\par}
\includegraphics[width=\subfigsize]{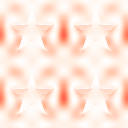}%
\hspace{1em}
\includegraphics[width=\subfigsize]{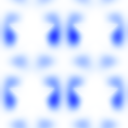}%
\par
}
\caption{\label{ener_dist_GaP_lattstar} 
Time-averaged distributions of the electric ($\varepsilon|\mathbf{E}|^2$; left panel, red color) and magnetic ($|\mathbf{H}|^2$; right panel, blue color) energy densities in the GaP metasurface with a square lattice of five-pointed star-shaped nanoholes at resonances ($h=100$~nm, $\lambda=532$~nm).
The geometric parameters of the lattices are displayed above.
The distributions are calculated within the plane $h/2$.
The incident Gaussian beam is polarized along the vertical direction.
The color scale is the same for all distributions.}
\end{figure}

The geometric parameters of the star-shaped nanohole lattice are varied in order to find resonances with strong electromagnetic energy confinement.
The distributions of the electric and magnetic energy densities are calculated for the lattice constants $a=240$, 250, 290, and 320~nm with the corresponding radii $r=104$, 122, 142, and 117~nm.
The time-averaged distributions of the electric $\varepsilon|\mathbf{E}|^2$ and magnetic $|\mathbf{H}|^2$ energy densities are shown in Fig.~\ref{ener_dist_GaP_lattstar}.
All distributions are calculated within the cross-sectional plane at $h/2$ and are presented on the same color scale.

For the lattices with $a=290$~nm, $r=142$~nm and $a=320$~nm, $r=117$~nm, a significant electric energy density is concentrated in air within the rays of the star-shaped holes.
The maximum electric energy density in these rays exceeds that in the Mie voids \cite{Panov26} by a factor of seven.
This makes the star-shaped nanohole lattices more promising than the metasurfaces with Mie voids for optical applications requiring field concentration.

\begin{figure}[tbh!]
{\centering
\includegraphics[width=\figsize]{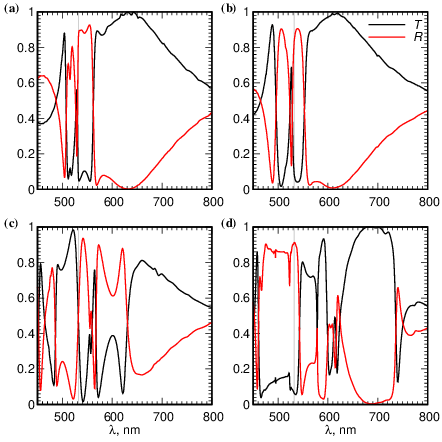}
\par} 
\caption{\label{Tspectr_GaP_lattstar}
Transmittance $T$ and reflectance $R$ spectra for the GaP metasurface with a square lattice of five-pointed star-shaped nanoholes ($h=100$~nm) with $a=240$~nm, $r=104$~nm (a); $a=250$~nm, $r=122$~nm (b); $a=290$~nm, $r=142$~nm (c); $a=320$~nm, $r=117$~nm (d).
The refractive index $n_{0\,\mathrm{in}}$ is assumed to be constant over the entire wavelength range.}
\end{figure}

The transmittance and reflectance spectra for the same geometric parameters are presented in Fig.~\ref{Tspectr_GaP_lattstar}.
For the lattices with $a=240$~nm, $r=104$~nm and $a=250$~nm, $r=122$~nm, the spectra are similar to those of metasurfaces with circular or polygonal holes exhibiting anapole states \cite{Panov22,Panov23}.

\begin{figure}[tbh!]
{\centering
\includegraphics[width=\figsize]{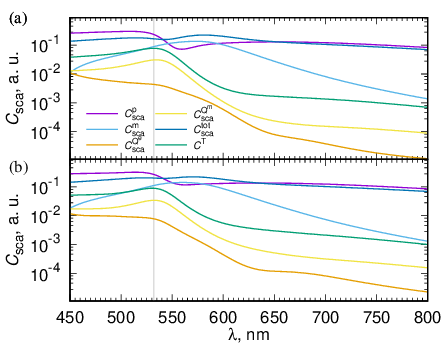}
\par} 
\caption{\label{Csca_GaP_lattstar}
Scattering cross section spectra for the multipole contributions (electric dipole $C_\mathrm{sca}^\mathrm{p}$, magnetic dipole $C_\mathrm{sca}^\mathrm{m}$, electric quadrupole $C_\mathrm{sca}^\mathrm{Q^e}$, magnetic quadrupole $C_\mathrm{sca}^\mathrm{Q^m}$), their sum $C_\mathrm{sca}^\mathrm{tot}$ and the intensity of the electric dipole toroidal moment $C^\mathrm{T}$ for the GaP lattice element outlined by the red line in Fig.~\ref{star_pores} with $h=100$~nm, $a=240$~nm, $r=104$~nm (a) and $a=250$~nm,  $r=122$~nm (b). 
Refractive index $n_{0\,\mathrm{in}}$ is assumed to be constant over the whole wavelength range.
}
\end{figure}

\begin{figure}[tbh!]
{\centering
\includegraphics[width=\figsize]{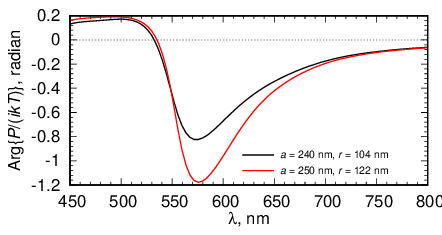}
\par} 
\caption{\label{Phi_del_star}
Phase difference between the electric dipole mode $P$ and the toroidal dipole mode $T$ for the GaP lattice element with $a=240$~nm, $r=104$~nm and $a=250$~nm, $r=122$~nm, $h=100$~nm, $k$ is wavenumber. 
The phase difference passes through zero near $\lambda=532$~nm, indicating the anapole state.
}
\end{figure}

The multipole decomposition of the scattering cross sections is presented in Fig.~\ref{Csca_GaP_lattstar} for the lattice elements outlined by the red square in Fig.~\ref{star_pores} with $a=240$~nm, $r=104$~nm and $a=250$~nm, $r=122$~nm.
Near $\lambda=532$~nm, a maximum of the electric dipole toroidal moment $C^{\mathrm{T}}$ is observed.
The phase difference between the electric dipole mode $P$ and the toroidal dipole mode $T$ is shown in Fig.~\ref{Phi_del_star}.
This phase difference passes through zero near $\lambda=532$~nm, which corresponds to the anapole state.
For the lattices with $a=290$~nm, $r=142$~nm and $a=320$~nm, $r=117$~nm, the multipole analysis does not reveal characteristic multipole peaks near this wavelength.

\begin{figure}[tbh!]
{\centering
\includegraphics[width=\figsize]{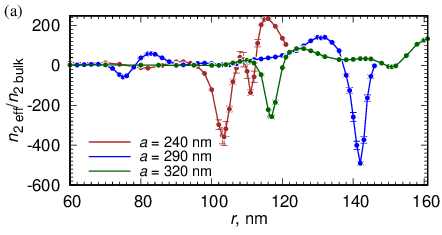}
\\
\includegraphics[width=\figsize]{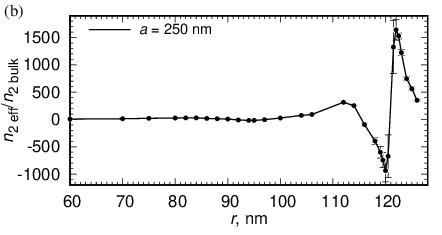}
\par} 
\caption{\label{n2_r_a_lattstar}
Enhancement of the effective second-order refractive index $n_{2{\mathrm{eff}}}/n_{2{\mathrm{bulk}}}$ for the GaP metasurface with a square lattice of five-pointed star-shaped nanoholes as a function of the star radius $r$ and the lattice constant $a$. 
The values of $a$ are displayed in the plot legend.}
\end{figure}

Next, the enhancement of the effective optical Kerr nonlinearity $n_{2{\mathrm{eff}}}/n_{2{\mathrm{bulk}}}$ is investigated as a function of the star radius $r$ for the lattice constants $a=240$, 250, 290, and 320~nm.
The results are presented in Fig.~\ref{n2_r_a_lattstar}.
For $a=240$~nm and $a=250$~nm, the dependencies exhibit sharp dips and peaks near the resonant radii $r=104$~nm and $r=122$~nm, respectively.
These resonances correspond to the anapole states identified by the multipole analysis.
The maximum value of $n_{2{\mathrm{eff}}}/n_{2{\mathrm{bulk}}}$ for $a=250$~nm, $r=122$~nm reaches approximately 1700, which has not been reported previously for GaP metasurfaces \cite{Panov22,Panov23}.
For $a=290$~nm and $a=320$~nm, the dependencies are smoother and the enhancement factors are lower, which is consistent with the absence of pronounced multipole resonances in these structures.

Near the resonances, the sign of $n_{2{\mathrm{eff}}}$ changes to negative.
The same effect was calculated previously for circular holes in the anapole state \cite{Panov22} and for spheres at the Mie resonances \cite{Panov19}.
A similar phenomenon was observed experimentally for the effective nonlinear dielectric function of metal nanoparticles in glass in the vicinity of a localized surface-plasmon resonance \cite{Takeda07,Takeda08,Sato14}.
In plasmonic nanocomposites, this effect is interpreted within the framework of the Maxwell-Garnett effective medium model as interference between the local and incident fields inside the nanoparticle.
Despite the considerably larger dimensions of all-dielectric meta-atoms, an analogous explanation may still apply.

\begin{figure}[tbh!]
{\centering
\includegraphics[width=\figsize]{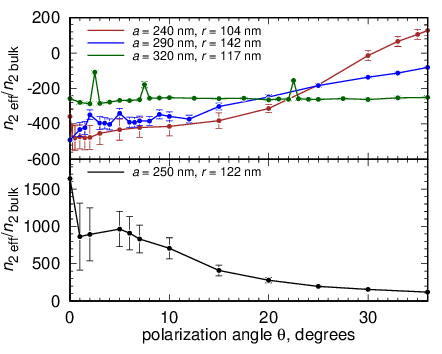}
\par} 
\caption{\label{n2_yrot_lattstar}
Enhancement of the effective second-order refractive index $n_{2{\mathrm{eff}}}/n_{2{\mathrm{bulk}}}$ for the GaP metasurface with a square lattice of five-pointed star-shaped nanoholes at $\lambda=532$~nm as a function of the polarization angle $\theta$ for the resonant geometric parameters. The values of $a$ and $r$ are displayed in the plot legend.
}
\end{figure}

Figure~\ref{n2_yrot_lattstar} illustrates the dependence of $n_{2{\mathrm{eff}}}/n_{2{\mathrm{bulk}}}$ on the polarization angle $\theta$ of the incident Gaussian beam.
The polarization angle is defined as the angle between the electric field vector $\mathbf{E}$ and the $x$-axis.
With increasing $\theta$, $|n_{2{\mathrm{eff}}}/n_{2{\mathrm{bulk}}}|$ generally decreases.
For $a=290$~nm, $r=142$~nm and $a=320$~nm, $r=117$~nm, $n_{2{\mathrm{eff}}}/n_{2{\mathrm{bulk}}}$ exhibits sharp changes at certain angles.
These changes are likely related to specific configurations of the field distribution and are unlikely to be observed in practice because of fabrication tolerances.

\begin{figure}[tbh!]
{\centering
\includegraphics[width=\figsize]{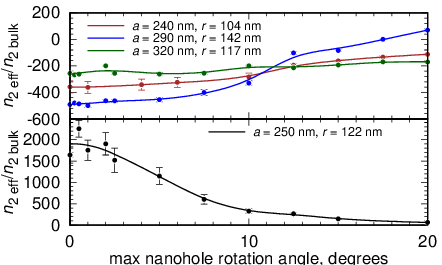}
\par} 
\caption{\label{n2_ra_lattstar}
Enhancement of the effective second-order refractive index $n_{2{\mathrm{eff}}}/n_{2{\mathrm{bulk}}}$ for the GaP metasurface with a square lattice of five-pointed star-shaped nanoholes at $\lambda=532$~nm as a function of the maximum random rotation angle for the resonant geometric parameters. The values of $a$ and $r$ are displayed in the plot legend.
}
\end{figure}

The influence of fabrication imperfections is modeled by introducing random rotation angles for each nanohole about its center in the membrane plane.
These random rotation angles are distributed uniformly in both clockwise and counterclockwise directions and do not exceed a specified maximum value.
The results are displayed in Fig.~\ref{n2_ra_lattstar}.
Random errors have only a weak effect on $n_{2{\mathrm{eff}}}/n_{2{\mathrm{bulk}}}$, gradually reducing its absolute value.
This demonstrates the robustness of the proposed metasurface against fabrication tolerances.

The star-shaped nanohole metasurfaces offer several advantages over other types of dielectric metasurfaces.
The high electric energy density concentrated in air within the rays of the star-shaped holes provides substantial field confinement, which is essential for enhancing nonlinear optical effects.
At the anapole state, the highest enhancement of the optical nonlinearity is achieved, exceeding the values reported previously for similar structures.
The robustness against fabrication imperfections and polarization variations makes these metasurfaces practical for real-world applications.
Taken together, these properties make the star-shaped nanohole metasurfaces a competitive platform for nonlinear optics.

In summary, the FDTD simulations demonstrate that a freestanding GaP membrane perforated with a square lattice of five-pointed star-shaped pores supports resonances with strong electromagnetic energy confinement and a pronounced enhancement of the optical Kerr effect.
The anapole states found for the lattices with $a=240$~nm, $r=104$~nm and $a=250$~nm, $r=122$~nm are accompanied by a sign reversal of the effective second-order refractive index and by enhancement factors reaching approximately 1700 with respect to bulk GaP, a value not previously reported for GaP metasurfaces.
For the lattices with $a=290$~nm, $r=142$~nm and $a=320$~nm, $r=117$~nm, the electric energy density localized in air within the rays of the star-shaped pores exceeds that of Mie voids by a factor of seven, although the nonlinear enhancement in these structures is lower and smoother.
The effective nonlinearity enhancement is shown to be robust against polarization rotation and random rotations of the nanoholes, which supports the feasibility of the proposed geometry under realistic fabrication tolerances.
The combination of strong field confinement in air, large nonlinear enhancement, and structural stability makes star-shaped pore membrane metasurfaces attractive for all-optical switching, nonlinear frequency conversion, and sensing applications.

\paragraph{Acknowledgement.} {\footnotesize
The research was carried out within the state assignment of IACP FEB RAS (Theme FWFW-2026-0006).
The results were obtained with the use of IACP FEB RAS Shared Resource Center ``Far Eastern Computing Resource'' equipment (https://www.cc.dvo.ru).}

\paragraph{Disclosures.} {\footnotesize
The author declares no conflicts of interest.}


\paragraph{Data availability.} {\footnotesize
Data underlying the results presented in this paper are not publicly available at this time but may be obtained from the authors upon reasonable request.}

\bibliography{nlphase}


\end{document}